\documentclass[10pt, pre, twocolumn, showpacs, nofootinbib]{revtex4-1}
\usepackage{url,ulem}
\usepackage{fourier}
\usepackage{amssymb,amsmath}
\usepackage[dvipdfmx]{graphicx}
\usepackage{graphicx,color}

\newcommand{\dfracp}[2]{\dfrac{\partial #1}{\partial #2}}

\newcommand{\norm}[1]{\left\Vert #1 \right\Vert}

\begin{document}

\title{Collective $1/f$ fluctuation by pseudo-Casimir-invariants}
\author{Yoshiyuki Y. Yamaguchi}
\email{yyama@amp.i.kyoto-u.ac.jp}
\affiliation{
  Department of Applied Mathematics and Physics, 
  Graduate School of Informatics, Kyoto University, 
  Kyoto 606-8501, Japan}
\author{Kunihiko Kaneko}
\affiliation{
  Department of Basic Science,
  University of Tokyo,
  3-8-1 Komaba, Meguro-ku, Tokyo 153-8902, Japan}

\begin{abstract}
  In this study, we propose a universal scenario
  explaining the $1/f$ fluctuation, including pink noises,
  in Hamiltonian dynamical systems with many degrees of freedom
  under long-range interaction. In the thermodynamic limit,
  the dynamics of such systems can be
  described by the Vlasov equation,
  which has an infinite number of Casimir invariants.
  In a finite system, they become pseudoinvariants,
  which yield quasistationary states.
  The dynamics then exhibit slow motion over them,
  up to the timescale where the pseudo-Casimir-invariants are effective.
  Such long-time correlation leads to $1/f$ fluctuations
  of collective variables,
  as is confirmed by direct numerical simulations.
  The universality of this collective $1/f$ fluctuation is demonstrated
  by taking a variety of Hamiltonians and changing the range of interaction
  and number of particles. 
\end{abstract}

\maketitle 

{\it Introduction:}
The $1/f$ fluctuation is ubiquitous in nature:
Certain quantities that fluctuate in time exhibit
a power spectrum of the form $1/f^{\nu}$,
where $f$ is the frequency and the exponent $\nu$ is
typically in the range of $1/2\lesssim\nu\lesssim 3/2$.
The $1/f$ fluctuation, suggesting a long-time correlation,
is observed in nature
such as
vacuum tubes \cite{johnson-25},
semiconductors \cite{caloyannides-74},
spin transport \cite{omar-etal-17},
oceans \cite{wunsch-72,taft-hickey-wunsch-baker-74},
quasars \cite{hawkins-01},
solar wind \cite{matthaeus-goldstein-86,matthaeus-etal-07},
and proteins \cite{min-etal-05}.
The $1/f$ fluctuation is also observed in model systems of
water molecules \cite{sasai-ohmine-ramaswamy-92,yamamoto-akimoto-yasui-yasuoka-15},
proteins \cite{yamamoto-akimoto-hirano-yasui-yasuoka-14},
ferromagnetic bodies \cite{yamaguchi-16},
and accretion disks \cite{janiuk-misra-12}.
(See, for instance, 
\cite{press-78,dutta-horn-81,weissman-88,milotti-02,ward-greenwood-07}
for reviews.)

There have been multiple efforts to understand
the mechanism of the $1/f$ fluctuation; however,
a coherent explanation remains to be lacking.
For instance,
the superposition of Lorentzians
(see \cite{milotti-02,ward-greenwood-07}, for instance)
requires a certain distribution of multiple timescales,
but the origin of multiple timescales must be explained.
Our strategy here is to restrict our concern
to Hamiltonian dynamical systems with many degrees of freedom
and search for the possibility of $1/f$ spectra for collective variables,
as observed in water molecules \cite{sasai-ohmine-ramaswamy-92}
and ferromagnetic bodies \cite{yamaguchi-16}.

In Hamiltonian systems with a few degrees of freedom,
the $1/f$ fluctuation has been observed
and analyzed by the hierarchical structure of the phase space constructed
by the KAM tori and chaotic sea
\cite{aizawa-84,meiss-85,meiss-86,meiss-ott-86,geisel-zacherl-radons-87,aizawa-kikuchi-harayama-yamamoto-ota-tanaka-89,yamaguchi-konishi-98}.
The hierarchical structure is analyzed by perturbation theory
in two-degrees-of-freedom systems \cite{lichtenberg-lieberman}
and is also observed in a symplectic coupled map with $4$ particles
\cite{konishi-kaneko-92}.
Nevertheless, such microscopic hierarchical structures exist
only in a certain range for a system with a few degrees of freedom
and is thus not generic in systems with many degrees of freedom.
Hence, it remains to be elucidated
how the $1/f$ fluctuation is generated in the collective motion
of many-degree-of-freedom systems.

The aim of this article is to propose a universal
scenario for the collective $1/f$ fluctuation
in long-range Hamiltonian systems with many degrees of freedom.
The target class of systems includes self-gravitating systems, plasmas, 
geophysical flows, 
trapped ions \cite{richerme-etal-14},
among others
\cite{campa-dauxois-ruffo-09,levin-etal-14,campa-dauxois-fanelli-ruffo}.
In the thermodynamic limit with an infinite number of particles,
the dynamics of such systems can be described
by the Vlasov equation, the partial differential equation
for the one-particle distribution function
\cite{braun-hepp-77,dobrushin-79,spohn-91}.
This equation is described by
the distribution function on the one-particle
phase space; therefore, the collective motion is naturally treated
through the Vlasov equation.

An important feature of Vlasov dynamics is that they have an infinite
number of Casimir invariants.
The Casimir-invariants are exact in the thermodynamic limit,
but in a system with a finite number of particles,
they become pseudoinvariants and fluctuate slowly with time.
These pseudo-Casimir-invariants
play the role of constraints up to a certain timescale,
but they break down in the long timescale,
as has also been observed recently
in the two-step relaxation of fluctuation amplitude
in thermal equilibrium 
\cite{yamaguchi-16}.
Such slow motion is owing to the pseudo-Casimir-invariants
and one may expect a long time correlation
in the dynamics of collective variables,
i.e., the collective $1/f$ fluctuation.
In the present article, we numerically demonstrate that this is indeed true
and propose a general scenario for the collective $1/f$ fluctuation
that persists  up to the timescale
where the constraint by pseudo-Casimir-invariants is effective.

{\it Model:}
We consider the $\alpha$-Hamiltonian mean-field ($\alpha$-HMF) model
\cite{anteneodo-tsallis-98},
which is described by the Hamiltonian
\begin{equation}
  \label{eq:Hamiltonian}
  H_{\alpha}(q,p) = \sum_{j=0}^{N-1} \dfrac{p_{j}^{2}}{2}
  + \dfrac{1}{2N_{\alpha}} \sum_{j=0}^{N-1} \sum_{k=0}^{N-1}
  \dfrac{1-\cos(q_{j}-q_{k})}{r_{jk}^{\alpha}},
\end{equation}
where $\alpha\geq 0$.
This system represents XY spins, each of which is located
at a one-dimensional lattice point with a unit lattice spacing.
The variable $q_{j}$ denotes the phase of the $j$th particle,
and $p_{j}$ is the conjugate momentum.
A quantum version of this system can be experimentally realized through
trapped ions \cite{richerme-etal-14}. Here,
the spatial boundary condition is set to be periodic,
and accordingly,
the distance between the $j$th and $k$th particles is defined as
$r_{jk}=\min\{~ |j-k|, ~ N-|j-k| ~ \}$ for $k\neq j$
and $r_{jk}=1$ for $k=j$.
The normalization factor $N_{\alpha}$ is introduced
to ensure the extensivity of energy, as is defined
by $N_{\alpha}=\sum_{k=0}^{N-1} 1/r_{jk}^{\alpha}$.
By taking $\alpha=0$ with $N_{0}=N$, the $\alpha$-HMF model
is reduced to the Hamiltonian mean-field (HMF) model
\cite{inagaki-konishi-93,antoni-ruffo-95}.
In the opposite limit, $\alpha\to\infty$, it can be reduced to the model
with the nearest-neighbor couplings with $N_{\infty}=3$.
Hence, the dependence on the coupling ranging from long to short
is investigated by varying the value of $\alpha$.

Here, we investigate the order parameter defined by
$\boldsymbol{M} = \sum_{j=0}^{N-1} (\cos q_{j}, \sin q_{j})/N$.
The $\alpha$-HMF model shows the second-order phase transition
at the specific energy $E_{\rm c}=3/4$,
which corresponds to the critical temperature $T_{\rm c}=1/2$,
irrespective of the value of $\alpha$ for $0\leq\alpha<1$
\cite{tamarit-anteneodo-00, campa-giansanti-moroni-00, mori-10}.
The boundary between the long- and short-range is given by $\alpha=1$,
beyond which the mean-field description is not valid.
We introduce the scaling $x=j/N$
such that the domain of $x$ is restricted to $x\in [-1/2,1/2]$
because of the periodic boundary condition.

In the limit $N\to\infty$,
the dynamics of the $\alpha$-HMF model
are described by the Vlasov equation \cite{bachelard-etal-11} :
\begin{equation}
  \label{eq:Vlasov}
  \dfracp{F}{t} + \dfracp{\mathcal{H}[F]}{p} \dfracp{F}{q}
  - \dfracp{\mathcal{H}[F]}{q} \dfracp{F}{p} = 0,
\end{equation}
where $F(q,p,x,t)$ is the one-particle distribution function,
$\mathcal{H}[F](q,p,x,t)$ is the one-particle Hamiltonian functional
defined by
\begin{equation}
  \label{eq:one-particle-Hamiltonian}
  \begin{split}
    & \mathcal{H}[F] = \dfrac{p^{2}}{2} + \mathcal{V}[F](q,x,t), \\
    & \mathcal{V}[F] = \dfrac{-1}{\kappa_{\alpha}}
    \int_{-1/2}^{1/2} dx' \int_{-\pi}^{\pi} dq' \int_{-\infty}^{\infty} dp' 
    \dfrac{\cos(q-q')}{|x-x'|^{\alpha}} F(q',p',x',t),
  \end{split}
\end{equation}
and $\kappa_{\alpha} = \int_{-1/2}^{1/2} \left( 1\left/ |x|^{\alpha} \right. \right) dx$.
From the Poisson structure of the Vlasov equation,
it is easy to show that a Casimir functional,
\begin{equation}
  \mathcal{C}[F](t)
  = \int_{-1/2}^{1/2} dx \int_{-\pi}^{\pi} dq \int_{-\infty}^{\infty} dp 
  ~ c(F(q,p,x,t))
\end{equation}
is a constant of motion for any differentiable function $c$
if $c(F)\to 0$ in $|p|\to\infty$ 
(see the Appendix \ref{sec:invariance-Casimir})
\cite{comment:cf}.
We stress that the validity of the Vlasov description is guaranteed
for $\alpha<1$, as the integral of $\kappa_{\alpha}$ does not converge
for $\alpha\geq 1$ \cite{comment:kappa-alpha}.

{\it Numerical tests and results:}
The initial values of $\{(q_{k},p_{k})\}_{k=0}^{N-1}$ 
are randomly chosen from 
the one-particle distribution function in thermal equilibrium:
\begin{equation}
  \label{eq:thermal-equilibrium}
  F_{\rm eq}(q,p; T,M) = A \exp\left[ -\left( p^{2}/2 - M\cos q \right)/T \right],
\end{equation}
where $A$ is the normalization factor, and $T$ is temperature.
From the rotational symmetry of the system, the direction
of the order parameter is set to $\boldsymbol{M}=(M,0)$.
The magnetization $M$ is determined
for a given $T$ by solving the self-consistent equation:
$M = \int_{-\pi}^{\pi} dq \int_{-\infty}^{\infty} dp ~ F_{\rm eq}(q,p; T, M) \cos q$.

The temperature $T$ is introduced to parameterize
the set of thermal equilibria.
All numerical simulations are performed by integrating
the canonical equations of motion associated with
the $N$-body Hamiltonian \eqref{eq:Hamiltonian} \cite{comment:FFT} 
without thermal noise
by using the fourth-order symplectic integrator
\cite{yoshida-93} with the time step $\Delta t=0.1$.

We recall that the Casimir-invariants hold under conditions of
(i) thermodynamic limit $N\to\infty$ and
(ii) long-range interaction ($\alpha<1$).
For large but finite $N$, the Casimir-invariants are no longer invariants;
they fluctuate with time, thus becoming pseudoinvariants.
To unveil the possible relationship between the pseudo-Casimir-invariants
and $1/f$ fluctuation,
we first numerically examine the $N$ dependence and $\alpha$ dependence 
in the low- and high-energy sides of the $\alpha$-HMF model, respectively.

\begin{figure}
  \centering
  \includegraphics[width=8cm]{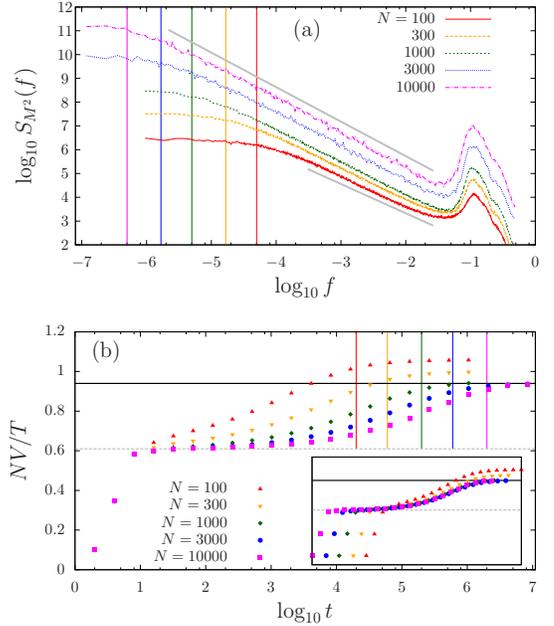}
  \caption{(color online)
    $N$ dependence in the HMF model ($\alpha=0$).
    $N=100$ ($300$, red triangles),
    $300$ ($300$, orange inverse triangles),
    $1000$ ($300$, green diamonds),
    $3000$ ($50$, blue circles), and $10000$ ($50$, magenta squares).
    The number in parentheses represents the number of sample orbits
    over which the average is taken. $T=0.45 (<T_{\rm c})$.
    (a) Power spectra of $M^{2}(t)$.
    $N$ increases from bottom to top.
    For graphical reasons, the vertical scales have been changed suitably.
    The vertical lines indicate the timescale $1/\tau$ where $\tau=200N$.
    The gray line segments guide the eyes for the slopes $-1.45$ (upper)
    and $-1.23$ (lower) obtained by the least-square method
    in the intervals of segments.
    (b) Temporal evolution of the scaled variance of magnetization $M$.
    The lower gray broken and upper black solid horizontal lines
    are, respectively, the theoretically predicted plateau level
    and thermal equilibrium level.
    $N$ increases from top to bottom.
    The vertical lines represent $\tau=200N$.
    (inset) The horizontal axis is $\log_{10}(t/N)$.}
  \label{fig:N-dependence-large}
\end{figure}

In the low-energy ordered phase $(T<T_{\rm c})$,
the relationship is examined through the $N$ dependence
by fixing the parameter $\alpha$ at $\alpha=0$ (the HMF model) for simplicity.
The power spectra of $M^{2}(t)$ are presented
in Fig.~\ref{fig:N-dependence-large}(a)
for several values of $N$, which exhibits $1/f$ fluctuations
down to a certain frequency $\tau^{-1}$.
From the figure, this maximum timescale is estimated as $\tau \sim 200N$,
suggesting that the $1/f$ fluctuation persists to small frequencies,
in the larger $N$.
The timescale $\tau \propto N$ is consistent with the fact
that the collision term of the product of two $O(1/\sqrt{N})$ terms 
is added to the Vlasov equation \eqref{eq:Vlasov} for finite $N$,
to break the Casimir invariants.
The power spectra of individual $\cos q_{j}(t)$
do not exhibit $1/f$ spectra (see the Appendix \ref{sec:PSD-cosq}),
and the observed $1/f$ spectra are caused by collective motion.
For a crossover of $1/f$ fluctuation between large and small $N$,
see the Appendix \ref{sec:small-N}.

To quantitatively reveal the timescale where the constraint by
the pseudo-Casimir-invariants is effective, 
we also compute the time-dependent variance $V(t)$ defined by
\begin{equation}
  \label{eq:Vt}
  V(t) = \sum_{n=0}^{n_{0}-1} \dfrac{1}{n_{0}} \left[
    \dfrac{1}{t} \sum_{k=1}^{t} \norm{\boldsymbol{M}(nt+k)}^{2}
    - \left(
      \dfrac{1}{t} \sum_{k=1}^{t} \norm{\boldsymbol{M}(nt+k)}
    \right)^{2}
  \right]
\end{equation}
where $t=2,4,8\cdots$ and $n_{0}$ is a sufficiently large number.
The variance $V$ is scaled as $NV/T$, which represents the susceptibility at
the thermal equilibrium level, according to
the fluctuation-response relation \cite{yamaguchi-16}.
The temporal evolution of the scaled variance 
is shown in Fig.~\ref{fig:N-dependence-large}(b)
with indications of the timescale $\tau=200N$
around which the variance reaches the asymptotic level.

From Fig.~\ref{fig:N-dependence-large}(b),
we can understand that the constraint
by the pseudo-Casimir-invariants is effective
up to the timescale $\tau$: 
The small collision term is negligible in the short-time regime
and the fluctuation is restricted in the initial Casimir level set,
which is an iso-Casimir surface in function space,
up to the end of the plateau \cite{comment:Vt-initial}.
The plateau level is theoretically predicted
using linear response theory for the Vlasov dynamics
\cite{ogawa-yamaguchi-12,ogawa-yamaguchi-15}
under the conservation of the Casimir invariants \cite{yamaguchi-ogawa-15}.
With time, the collision term becomes non-negligible,
altering the Casimir invariants and moving to another level set.
The state is trapped on the new level set
(although the next plateau is invisible in the figure
because of the logarithmic axis).
The change in level sets continues and
the cumulative variance \eqref{eq:Vt} thus slowly increases,
whereas the suppression by the pseudo-Casimir-invariants remains effective.
We underline that the plateau is not perfectly flat,
where a perfect plateau suggests that the Casimir invariants are exact
as in the limit $N\to\infty$.
The arrival to the asymptotic level at $\tau$ suggests that the state 
has traveled over the possible level sets
and the constraint by the pseudo-Casimir-invariants is no longer effective
for timescales larger than $\tau$
\cite{comment:time-shift}.

\begin{figure}[h]
  \centering
  \includegraphics[width=12cm]{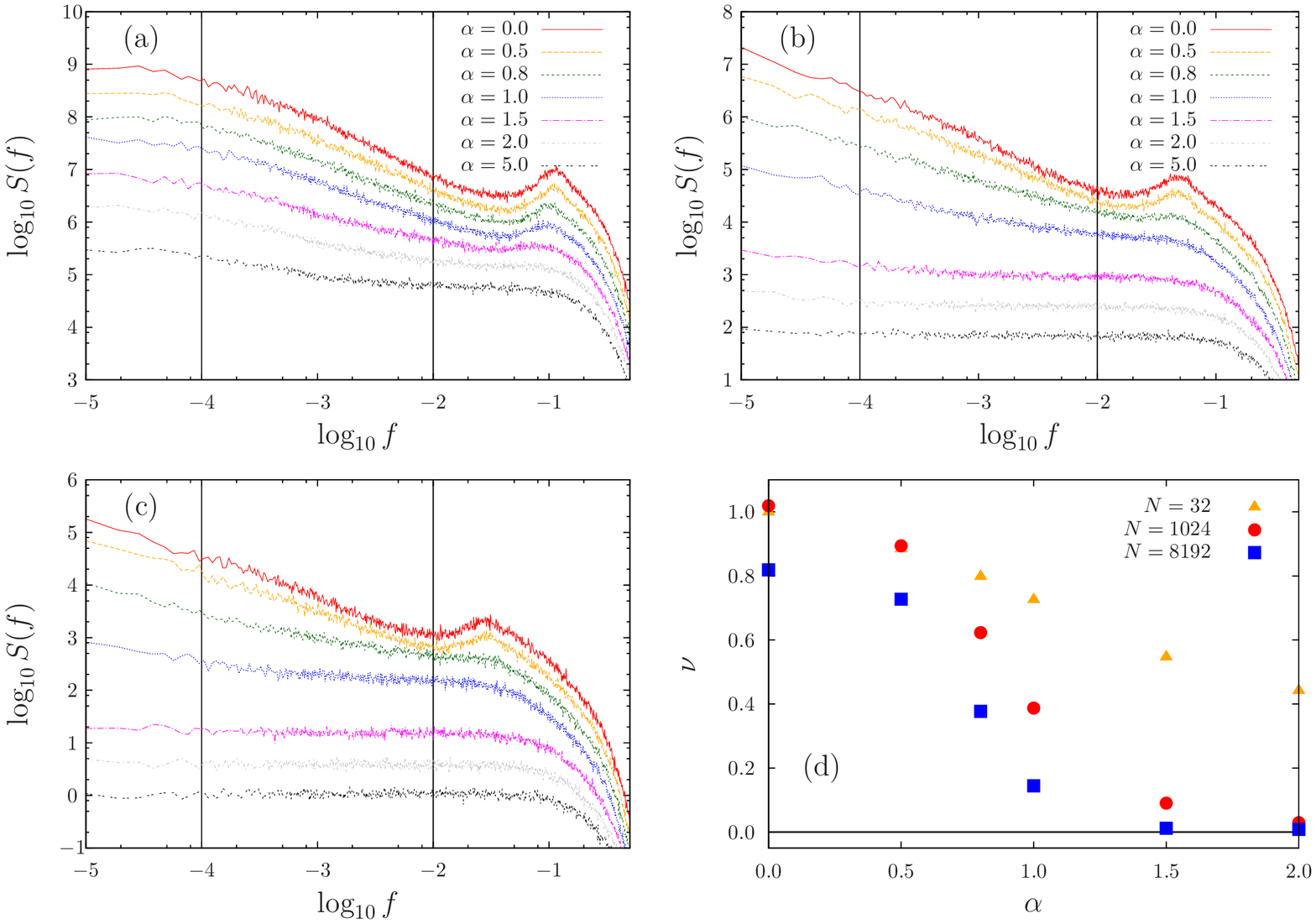}
  \caption{(color online)
    Power spectra in the $\alpha$-HMF model. $T=0.6$.
    (a) $N=32$ ($100$). (b) $N=1024$ ($100$). (c) $N=8192$ ($60$).
    The number in parentheses represents the number of sample orbits.
    $\alpha=0,~ 0.5,~ 0.8,~ 1,~ 1.5,~ 2$, and $5$ from top to bottom.
    For graphical reasons, the vertical scale has been modified suitably.
    (d) $\alpha$ dependence of the exponent $\nu$ (minus of the slope),
    which is computed by the least-square method
    between the two vertical lines of the panels (a)-(c).}
  \label{fig:alphaHMF}
\end{figure}

In the high-energy disordered phase $(T>T_{\rm c})$,
linear response theory predicts that the variance on a Casimir level set
agrees with the thermal equilibrium level
\cite{ogawa-yamaguchi-12,patelli-gupta-nardini-ruffo-12}, 
and, accordingly, no two-step relaxation of $V(t)$ appears
(see the Appendix \ref{sec:N-dep-high}).
In contrast, the disordered phase exists
for any value of $\alpha$.
We, therefore, change the strategy and
investigate the dependence of the range of interaction $\alpha$
by choosing the initial condition from thermal equilibrium
\eqref{eq:thermal-equilibrium} by setting $M=0$ for $T=0.6 (>T_{\rm c})$.
The power spectra of $M^{2}(t)$ for $T=0.6$ are presented
in Fig.~\ref{fig:alphaHMF} for $N=32, 1024$, and $8192$.
As we expected, the $1/f$ spectrum tends to disappear as $\alpha$ increases,
i.e., as the interaction range becomes shorter.
Moreover, this tendency is enhanced by increasing $N$
and the exponent $\nu$ goes to $0$ in $\alpha>1$
as reported in Fig.~\ref{fig:alphaHMF}(d).
Note that the collective motion is again necessary
for observing the $1/f$ spectrum
(see the Appendix \ref{sec:PSD-cosq}).
At the low-energy end, a similar figure with Fig.~\ref{fig:alphaHMF}
has the same tendency but does not coincide exactly
(see the Appendix \ref{sec:alpha-dep-low}).

\onecolumngrid

\vspace*{2em}

\newpage
\twocolumngrid

\begin{figure}
  \centering
   \includegraphics[width=6.4cm]{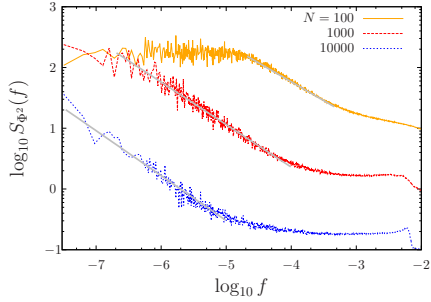}
  \caption{(color online)
    Power spectra of $\boldsymbol{\Phi}^{2}(t)$
    in the globally coupled FPUT model. 
    The initial state is in thermal equilibrium with $T=1$.
    $N=100$ (top orange), $1000$ (middle red), and $10000$ (bottom blue),
    where the vertical scales of the last two lines have been multiplied
    by $10$ and $100$, respectively, for graphical reasons.
    Each curve is the averages over $20$ samples.
    The time step of simulations is $\Delta t=0.01$.
    The initial interval $t\in [0,10^{4}]$ has been removed
    to avoid transience \cite{comment:InitialTransience}.
    The gray line segments are obtained from the least-square method
    in the intervals of segments.
    The slopes are $-0.62$, $-0.70$, and $-0.74$
    from top to bottom.
    }
  \label{fig:FPUT}
\end{figure}

Finally, we investigate whether the existence of phase transition
in the $\alpha$-HMF model is essential to the $1/f$ fluctuation.
To this end, we introduce a globally coupled 
Fermi-Pasta-Ulam-Tsingou (FPUT) system \cite{dauxois-08}:
\begin{equation}
  \label{eq:FPUT}
  H_{\rm FPUT} = \sum_{j=0}^{N-1} \dfrac{p_{j}^{2}}{2}
  + \dfrac{1}{2N} \sum_{j=0}^{N-1} \sum_{k=0}^{N-1} \varphi(q_{j}-q_{k}),
  \quad
  \varphi(q) = \dfrac{q^{2}}{2} + \dfrac{q^{4}}{4},
\end{equation}
which does not exhibit the phase transition. 
The collective observables that are na{\"i}vely expected,
such as the variance of $q$ or the potential energy, do not exhibit
the $1/f$ fluctuation
(see the Appendix \ref{sec:dep-observables}).
In contrast, the power spectrum of the collective variable
$\boldsymbol{\Phi}^{2}(t)$ exhibits $1/f$ fluctuation
as shown in Fig.~\ref{fig:FPUT},
where $\boldsymbol{\Phi}=\sum_{j}(\cos\phi_{j},\sin\phi_{j})/N$
and $\phi$ is defined on the $(q,p)$ plane
as $(q,p)=(r\cos\phi, r\sin\phi)$ with $r=\sqrt{q^{2}+p^{2}}$.
To understand this result, we note that on one Casimir level set,
fluctuations should occur on each iso-action curve
in the leading order \cite{yamaguchi-ogawa-15},
where the action variable is associated with
the one-particle Hamiltonian for the thermal equilibrium state.
Traveling over level sets gives rise fluctuations over iso-action curves,
whereas the angle variable, conjugate with the action variable,
captures the long-time correlation
(see the Appendix \ref{sec:dep-observables}).
The variable $\phi$ here corresponds to the angle variable
\cite{comment:action-angle}.
Therefore, $\boldsymbol{\Phi}$ is a suitable collective variable
to extract the hidden $1/f$ fluctuation arising from
the pseudo-Casimir-invariants.

{\it Summary and discussions:}
In this study, we investigated
the origin of the collective $1/f$ fluctuation
in many-degree-of-freedom Hamiltonian systems under long-range interaction.
Such systems have an infinite number of Casimir invariants
in the limit $N\to\infty$, which, in a finite system,
constrains the motion as pseudoinvariants
and leads to the slow motion of collective variables. 
We then propose a simple but universal scenario:
The collective $1/f$ fluctuation appears in the timescale up to which
the constraint by the pseudo-Casimir-invariants is effective.
This scenario has been successfully verified numerically 
by investigating the $\alpha$-HMF model, FPUT model,
and the dependence of the power spectra
on the number of elements and the interaction range.

As the Casimir-invariants are based on the distribution function
and the average of an observable over it gives a collective variable,
it is natural that the $1/f$ fluctuation is observed
only for collective variables.
For local variables such as the position
of each particle, such long-term fluctuations are not observed.
The collective variable in concern is given by an order parameter,
as the average of the angle variable, corresponding to the slow change
of the level of pseudo-Casimir-invariants.
In the globally coupled FPUT model,
the $1/f$ fluctuation may be hidden and is extracted by 
setting suitable observables with the aid of the proposed scenario.

To confirm the importance of the Casimir invariants
(that are based on the Poisson structure of the Vlasov equation),
we also examined the Kuramoto model \cite{kuramoto-75},
as a non-Hamiltonian and dissipative version of the HMF model,
and confirmed that the system does not exhibit $1/f$ fluctuation
(see the Appendix \ref{sec:kuramoto-model})
\cite{garcia-gudino-etal-17}.
The superposition of Lorentzian spectra is also unsuitable
to explain the observed $1/f$ fluctuation
by considering the timescales determined by the Landau damping modes
(see the Appendix \ref{sec:superposition-lorentzian}).

It is important to examine the universality of our result.
Although we considered the thermal equilibrium states
of simple models here,
the existence of Casimir invariants does not depend on
consideration of reference states or the
details of interaction potentials in 
long-range Hamiltonian systems,
and the slow dynamics by pseudo-Casimir-invariants will
generally be applied to
the so-called quasistationary states \cite{campa-dauxois-ruffo-09}
for instance.
As there exists a variety of examples that can be
effectively described by long-range Hamiltonian systems,
such as plasmas,
free electron laser
\cite{campa-dauxois-ruffo-09,colson-76,bonifacio-pellegrini-narducci-84,zaslavsky-etal-77,barre-etal-04},
water molecules \cite{jorgensen-81}, 
and trapped ions \cite{richerme-etal-14,porras-cirac-04,kim-etal-09,britton-etal-12,islam-etal-13},
collective $1/f^{\nu}$ fluctuation will be experimentally verifiable.
In real systems, the Poisson structure providing the Casimir invariants
will be disturbed by dissipation and randomness.
It will be important to examine the robustness of our scenario
under such perturbations.
Finally, the theory to predict the exponent $\nu$ in the $1/f^{\nu}$ spectrum
must be formulated based on the slow motion of pseudo-Casimir-invariants.

\acknowledgements
Y.Y.Y. acknowledges the support of
JSPS KAKENHI Grant No. 16K05472.

\appendix

\onecolumngrid

\section{Casimir invariants}
\label{sec:invariance-Casimir}

Let us consider the Vlasov equation
\begin{equation}
  \label{eq:Vlasov-eq}
  \dfracp{F}{t} + \dfracp{\mathcal{H}[F]}{p} \dfracp{F}{q}
  - \dfracp{\mathcal{H}[F]}{q} \dfracp{F}{p} = 0
\end{equation}
for the one-particle distribution function $F(q,p,x,t)$,
where $q$ and $p$ are the conjugate position and momentum,
and $x$ represents the spatial position.
We prove that a Casimir functional
\begin{equation}
  \mathcal{C}[F](t)
  = \int_{-1/2}^{1/2} dx \int_{-\pi}^{\pi} dq \int_{-\infty}^{\infty} dp 
  ~ c(F(q,p,x,t))
\end{equation}
is a constant of motion
if the one-particle Hamiltonian functional $\mathcal{H}[F](q,p,x,t)$
belongs to the $C^{2}$-class with respect to $q$ and $p$,
and if $|c(F(q,p,x,t))|$ rapidly decreases in $|p|\to\infty$.
Differentiating with respect to time $t$ and using the Vlasov equation, we have
\begin{equation}
  \begin{split}
    \dfrac{d\mathcal{C}[F]}{dt}
    & = \int_{-1/2}^{1/2} dx \int_{-\pi}^{\pi} dq \int_{-\infty}^{\infty} dp 
    ~ c'(F) \left( - \dfracp{\mathcal{H}[F]}{p} \dfracp{F}{q}
      + \dfracp{\mathcal{H}[F]}{q} \dfracp{F}{p} \right) \\
    & = \int_{-1/2}^{1/2} dx \int_{-\pi}^{\pi} dq \int_{-\infty}^{\infty} dp 
    ~ \left( - \dfracp{\mathcal{H}[F]}{p} \dfracp{c(F)}{q}
      + \dfracp{\mathcal{H}[F]}{q} \dfracp{c(F)}{p} \right). \\
  \end{split}
\end{equation}
Performing the integration by parts and noting the rapid decrease
in $c(F)$ for $|p|\to\infty$, we have
\begin{equation}
  \begin{split}
    \dfrac{d\mathcal{C}[F]}{dt}
    & = \int_{-1/2}^{1/2} dx \int_{-\pi}^{\pi} dq \int_{-\infty}^{\infty} dp 
    ~ c(F) \left( \dfracp{}{q} \dfracp{\mathcal{H}[F]}{p}
      - \dfracp{}{p} \dfracp{\mathcal{H}[F]}{q} \right)
    = 0, \\
  \end{split}
\end{equation}
for $\mathcal{H}[F]$ in the $C^{2}$-class.
The decreasing speed of $c(F)$ is sufficient to cause the term
$c(F) \partial_{q}\mathcal{H}[F]$ to vanish in the limit $|p|\to\infty$,
and the condition is reduced to $c(F)\to 0$ for $|p|\to\infty$
if $\mathcal{H}[F]$ consists of the $p$-dependent kinetic
and $q$-dependent potential parts.

\section{Power spectra of individual particles}
\label{sec:PSD-cosq}

\begin{figure}[h]
  \centering
  \includegraphics[width=8cm]{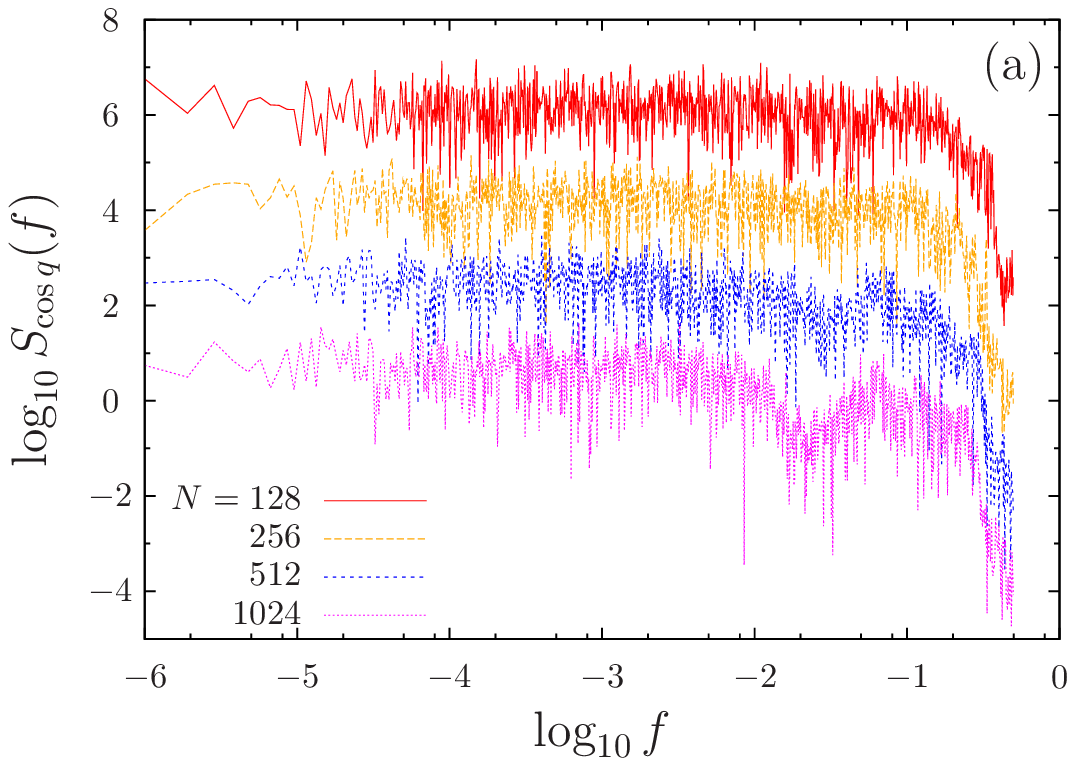}
  \includegraphics[width=8cm]{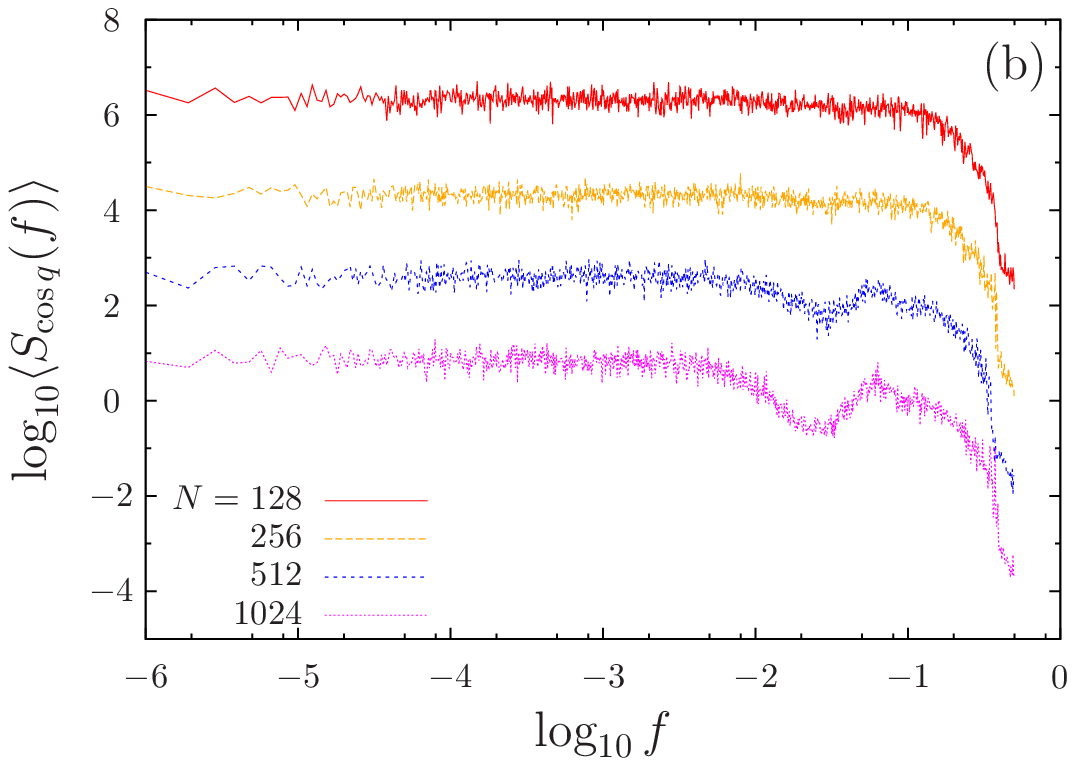}
  \caption{(color online)
    (a) Power spectra of $\cos q_{0}(t)$
    and (b) average over $N$ of power spectra of $\cos q_{j}(t)$
    in the $\alpha$-HMF model. $\alpha=0$. $T=0.45$.
    $N=128$ (red), $256$ (orange), $512$ (blue), and $1024$ (magenta)
    from top to bottom,
    where the horizontal scales have been multiplied by
    $10^{-2},~10^{-4}$, and $10^{-6}$ in the last three curves, respectively,
    for graphical reasons.}
  \label{fig:Power-alphaHMF-indivcosq-Ndep}
\end{figure}

\begin{figure}[h]
  \centering
  \includegraphics[width=8cm]{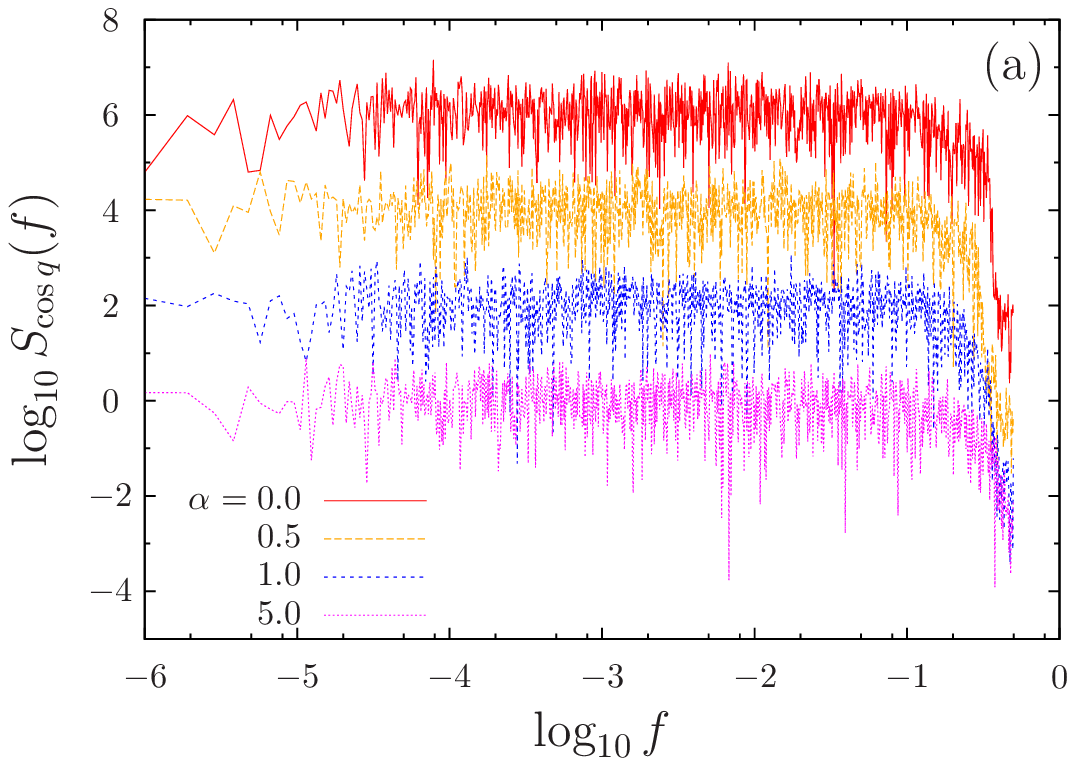}
  \includegraphics[width=8cm]{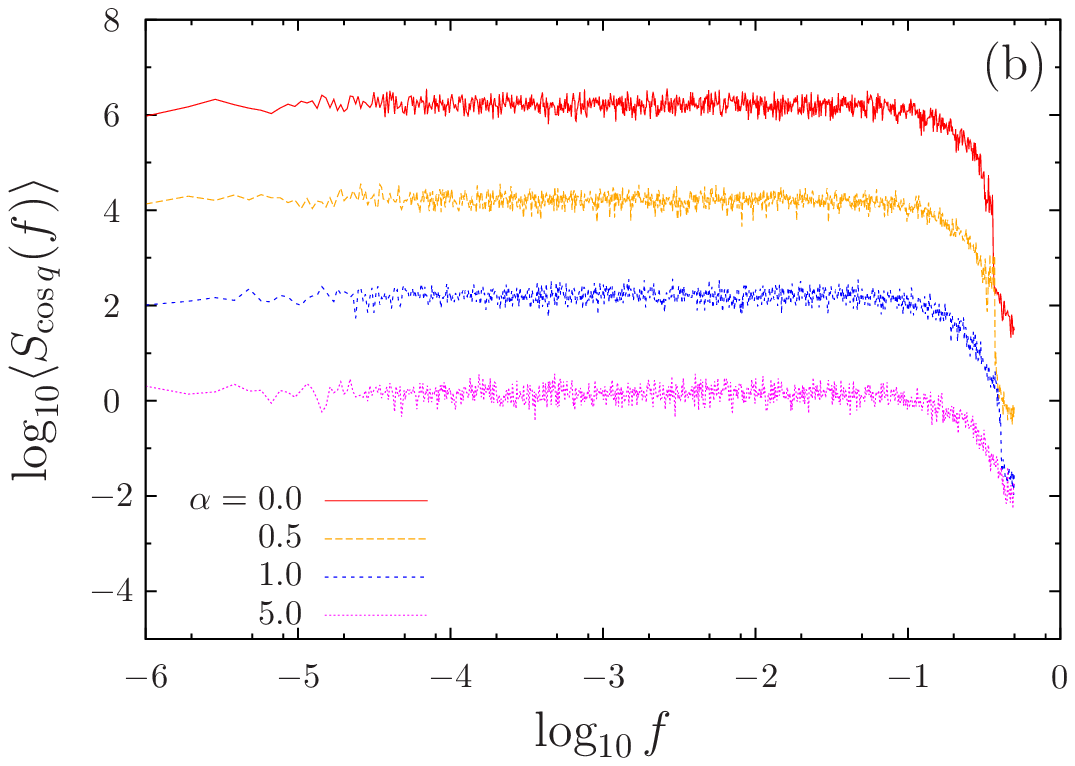}
  \caption{(color online)
    (a) Power spectra of $\cos q_{0}(t)$
    and (b) average over $N$ of power spectra of $\cos q_{j}(t)$
    in the $\alpha$-HMF model. $N=1024$. $T=0.6$.
    $\alpha=0$ (red), $0.5$ (orange), $1$ (blue), and $5$ (magenta)
    from top to bottom,
    where the horizontal scales have been multiplied by
    $10^{-2},~10^{-4}$, and $10^{-6}$ in the last three curves, respectively,
    for graphical reasons.}
  \label{fig:Power-alphaHMF-indivcosq-alphadep}
\end{figure}

In the $\alpha$-Hamiltonian mean-field model,
no $1/f$ fluctuation occurs in the time series
concerning the individual particle.
Figures \ref{fig:Power-alphaHMF-indivcosq-Ndep}
and \ref{fig:Power-alphaHMF-indivcosq-alphadep}
respectively report flat power spectra
for $N$ dependence with $\alpha=0$ and $\alpha$ dependence with $N=1024$.
We thus conclude that the observed $1/f$ fluctuation comes
from collective motion.

\section{$N$ dependence in low-energy ordered phase with small $N$}
\label{sec:small-N}

\begin{figure}
  \centering
  \includegraphics[width=8cm]{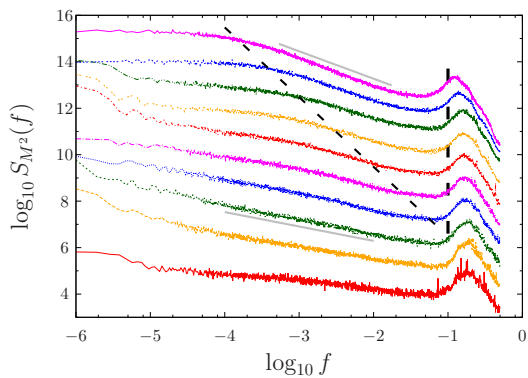}
  \caption{(color online)
    Power spectra of $M^{2}(t)$ in the HMF model ($\alpha=0$).
    $T=0.45 (<T_{\rm c})$.
    $N=3, 4, \cdots, 10, 20$, and $50$ from bottom to top.
    For graphical reasons, the vertical scale has been multiplied
    by $10^{1}$ for $N=4$, $10^{2}$ for $N=5$, etc.
    Each curve is the average over $100$ samples.
    The gray line segments guide the eyes for the slopes
    $-0.62$ (lower) and $-1.18$ (upper) obtained by the least-square method
      in the intervals of the segments.
    The $1/f$ fluctuation due to pseudo-Casimir-invariants
    is proposed as the enclosed domain by the two black dashed lines.
   }
  \label{fig:N-dependence-small}
\end{figure}

By further investigating the $N$ dependence of the power spectrum,
we note the existence of two distinct types of 
$1/f$ fluctuation as in Fig.~\ref{fig:N-dependence-small}.
For larger $N(\gtrapprox 9)$, the collective $1/f$ fluctuation is observed.
For smaller $N~(3\lessapprox N\lessapprox 8)$,
however, the $1/f$ fluctuation 
exhibits slopes clearly different from the previous type.
One possible explanation of the $1/f$ fluctuation with small $N$
might be the hierarchical structure of phase space
\cite{konishi-kaneko-92},
whereas we note that $10$ particles are
sufficiently large
to exhibit the collective $1/f$ fluctuation based on the
pseudo-Casimir-invariants.

\section{$N$ dependence in high-energy disordered phase}
\label{sec:N-dep-high}

\begin{figure}[h]
  \centering
  \includegraphics[width=8cm]{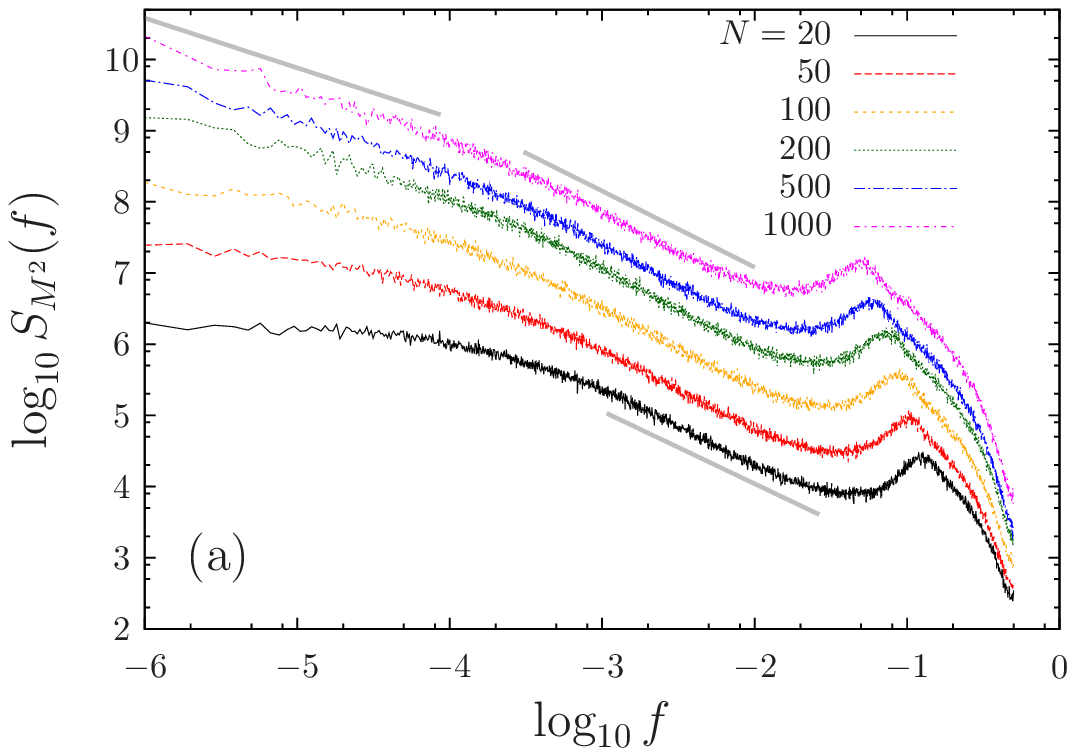}
  \includegraphics[width=8cm]{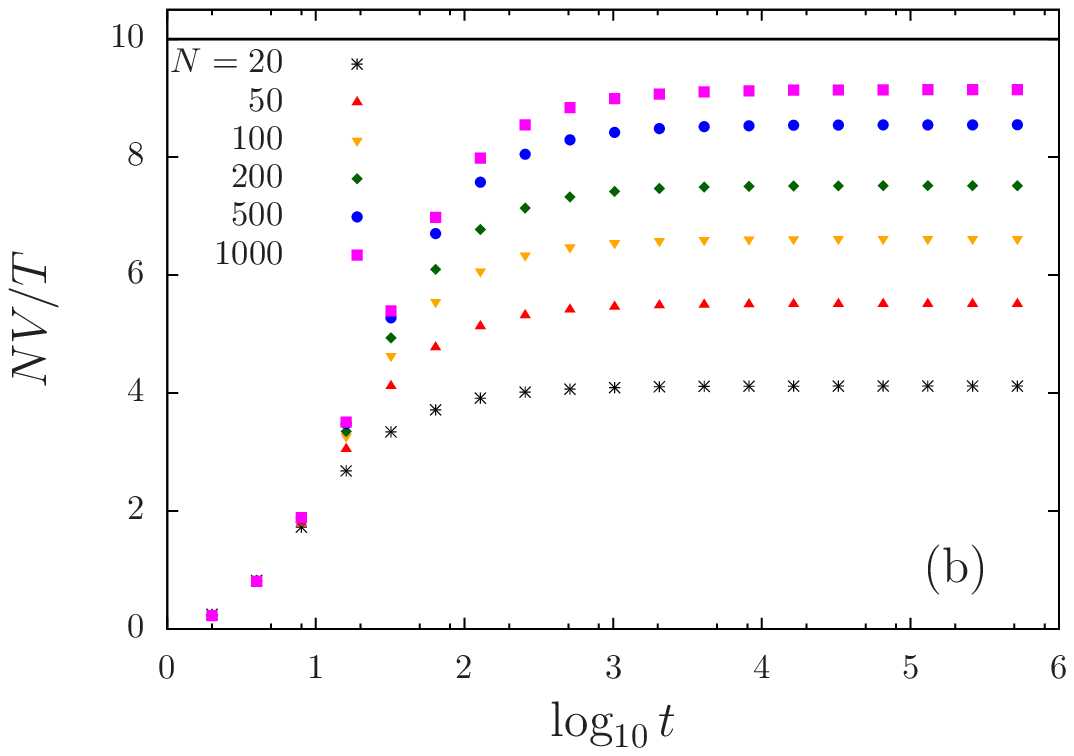}
  \caption{(color online)
    $N$ dependence in the HMF model ($\alpha=0$). $T=0.60 (>T_{\rm c})$.
    $N=20$ (black starts), $50$ (red triangles),
    $100$ (orange inverse triangles),
    $200$ (green diamonds),
    $500$ (blue circles), and $1000$ (magenta squares).
    (a) Power spectra of $M^{2}(t)$.
    $N$ increases from bottom to top.
    For graphical reasons, the vertical scale has been multiplied
    by $10^{1}$ for $N=50$, $10^{2}$ for $N=100$, etc.
    The gray line segments guide the eyes for the slopes
    $-0.70$ (left upper), $-1.07$ (right upper),
    and $-1.02$ (right lower) obtained by the least-square method
    in the intervals of segments.
    (b) Temporal evolution of the scaled variance of magnetization $M$.
    $N$ has the same values as those in (a).
    The black horizontal line represents the thermal equilibrium level, $10$.
    In both panels, each curve is the average over $100$ samples.}
  \label{fig:T0.60}
\end{figure}

\begin{figure}[h]
  \centering
  \includegraphics[width=8cm]{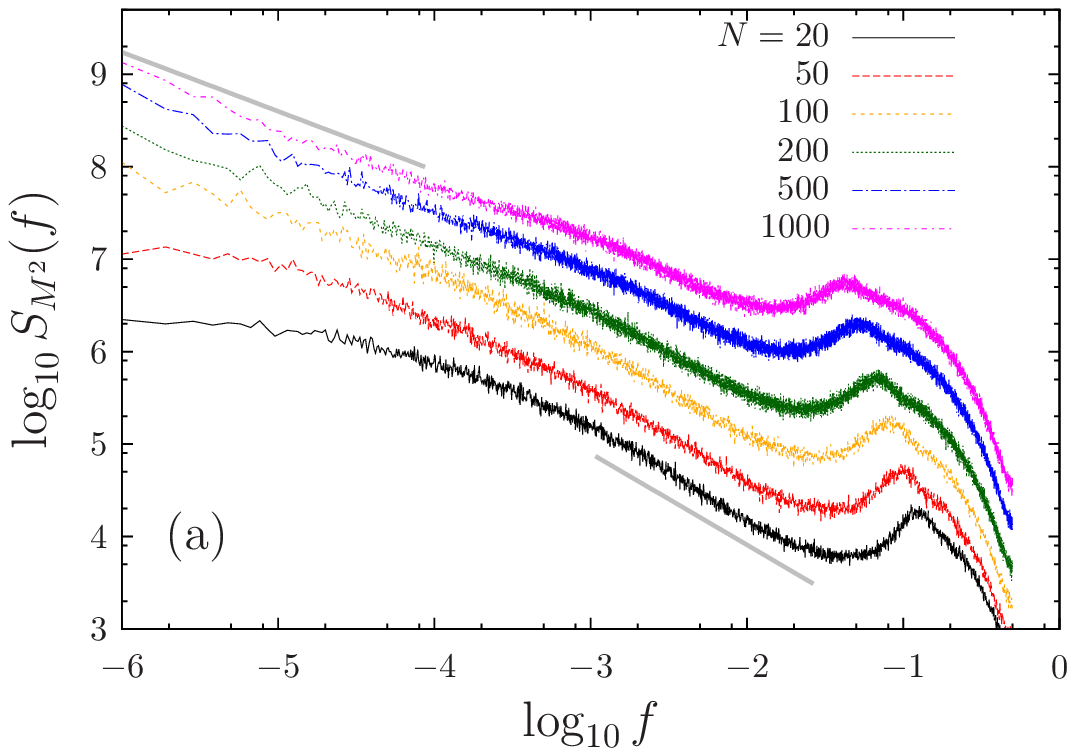}
  \includegraphics[width=8cm]{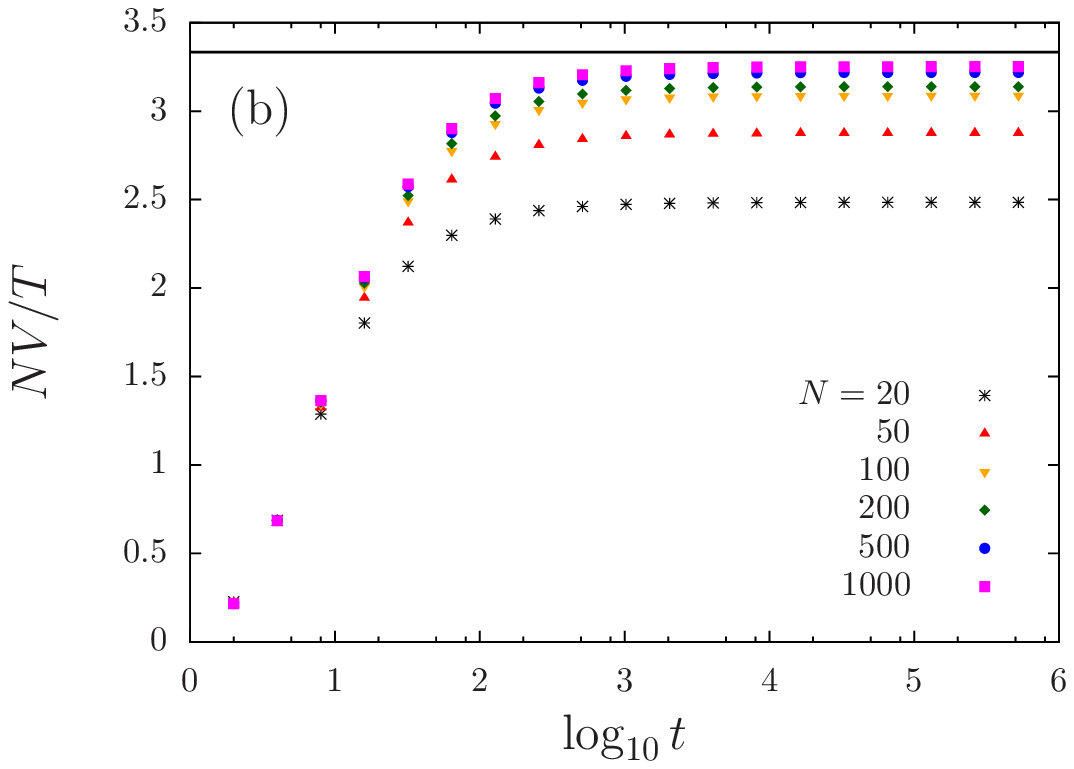}
  \caption{(color online)
    Similar to Fig.~\ref{fig:T0.60} but with $T=0.8$.
    In panel (a), the vertical scales are suitably modified.
    The gray line segments guide the eyes 
    for the slopes $-0.64$ (upper left) and $-0.99$ (lower right)
    obtained by the least-square method
    in the intervals of segments.
    In panel (b), the black horizontal line
    represents the thermal equilibrium level, $10/3$.}
  \label{fig:T0.80}
\end{figure}

Next, we study the power spectra of the square order parameter
and temporal evolution of the variance
in the high-energy disordered phase
of the Hamiltonian mean-field model $(\alpha=0)$,
by setting the temperature
to $T=0.6$ in Fig.~\ref{fig:T0.60}
and $T=0.8$ in Fig.~\ref{fig:T0.80},
where the critical temperature is $T_{\rm c}=0.5$.
At both temperatures, $1/f$ power spectra are observed, but
the slopes $-0.70$ and $-0.64$ in the long-time region are quite close
to that observed with small $N$ in the low-energy ordered phase.
In the high-energy side, the largest Lyapunov exponent
tends to vanish as $N$ increases
\cite{yamaguchi-96,latora-rapisarda-ruffo-98,firpo-98}.
Accordingly, one possible understanding of the small exponent is
that the system is nearly integrable
and the hierarchical structure appears
together with the constraint by the pseudo-Casimir-invariants.
The comparison between $T=0.6$ and $T=0.8$ supports this
hypothesis as the small exponent becomes significant
as $T$ increases, where the dynamics are more regular.

At the high-energy side, we define the variance as
\begin{equation}
  V(t) = \sum_{n=0}^{n_{0}-1} \dfrac{1}{n_{0}} \left[
    \dfrac{1}{t} \sum_{k=1}^{t} \norm{\boldsymbol{M}(nt+k)}^{2}
    - \left(
      \dfrac{1}{t} \sum_{k=1}^{t} \boldsymbol{M}(nt+k)
    \right)^{2}
  \right],
\end{equation}
because the order parameter vector $\boldsymbol{M}$ fluctuates 
on the two-dimensional plane around $\boldsymbol{M}={\bf 0}$.
Both with and without the Casimir constraints,
the scaled variance $NV/T$ is theoretically predicted
as $NV/T=2T_{\rm c}/(T-T_{\rm c})$
\cite{yamaguchi-16},
which is twice the susceptibility.
Accordingly, there is no two-step relaxation of variance
in Figs.~\ref{fig:T0.60}(b) and \ref{fig:T0.80}(b).
The arrival to the thermal equilibrium level, therefore, does not imply
the end of effectivity of the pseudo-Casimir-invariants,
because the arrival is possible on one Casimir level set.

\section{$\alpha$ dependence in low-energy ordered phase}
\label{sec:alpha-dep-low}

\begin{figure}[h]
  \centering
  \includegraphics[width=14cm]{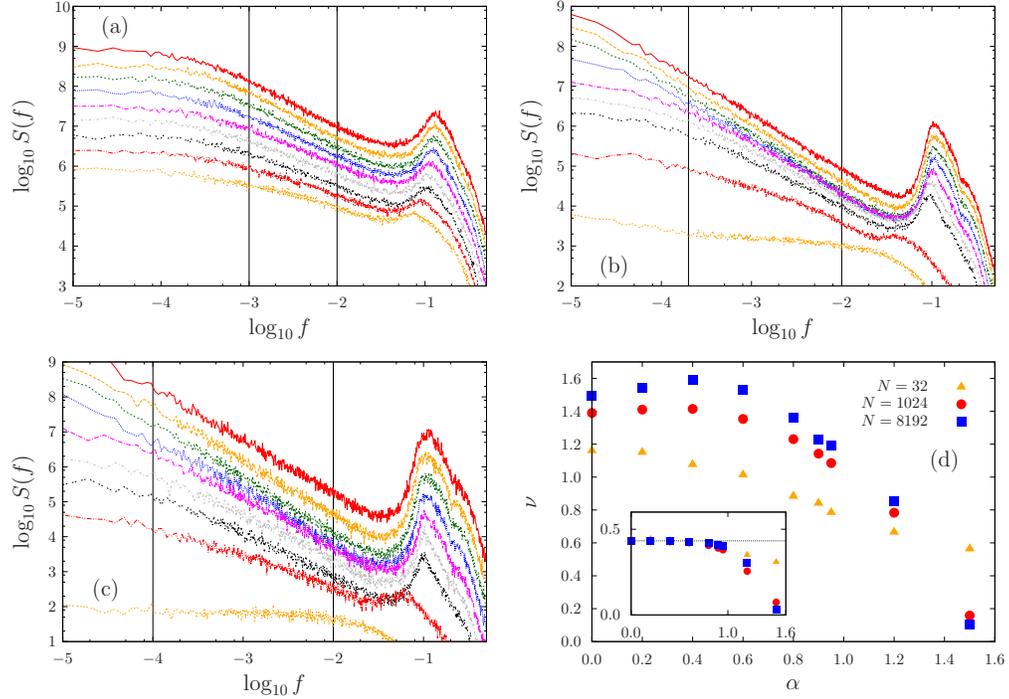}
  \caption{(color online)
    Power spectra in the $\alpha$-HMF model. $T=0.45$.
    (a) $N=32$ ($100$). (b) $N=1024$ ($100$). (c) $N=8192$ ($20$).
    The number in parentheses represents the number of sample orbits.
    $\alpha=0$ (red),
    $0.2$ (orange),
    $0.4$ (green),
    $0.6$ (blue),
    $0.8$ (magenta),
    $0.9$ (gray),
    $0.95$ (black),
    $1.2$ (red),
    and $1.5$ (orange)
    from top to bottom.
    For graphical reasons, the vertical scale has been modified suitably.
    (d) $\alpha$ dependence of the exponent $\nu$ (minus of the slope),
    which is computed by the least-square method
    between the two vertical lines of the panels (a)-(c).
    $N=32$ (orange triangles), $1024$ (red circles), and $8192$ (blue squares).
    (inset) $\alpha$ dependence of time average of $M$.}
  \label{fig:alphaHMF-low}
\end{figure}

\begin{figure}[h]
  \centering
  \includegraphics[width=14cm]{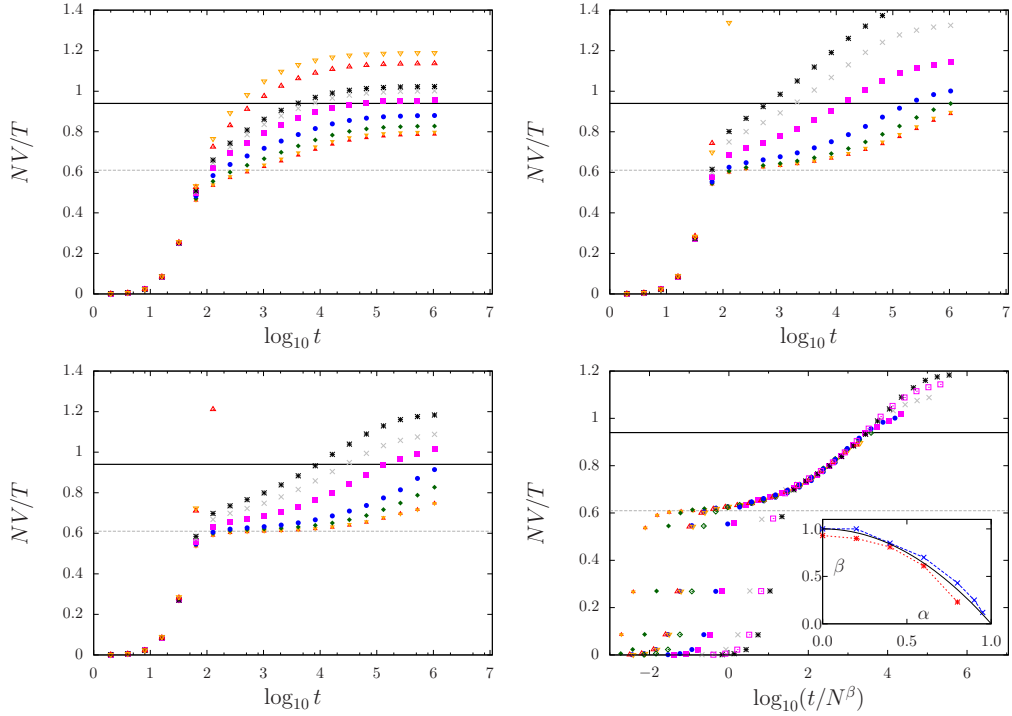}
  \caption{(color online)
    Temporal evolution of $V(t)$ in the $\alpha$-HMF model. $T=0.45$.
    (a) $N=32$ ($100$). (b) $N=1024$ ($100$). (c) $N=8192$ ($20$).
    The number in parentheses represents the number of sample orbits.
    $\alpha=0$ (red triangles),
    $0.2$ (orange inverse triangles),
    $0.4$ (green diamonds),
    $0.6$ (blue circles),
    $0.8$ (magenta squares),
    $0.9$ (gray crosses),
    $0.95$ (black stars),
    $1.2$ (red open triangles),
    and $1.5$ (orange open inverse triangles)
    from bottom to top.
    The horizontal scales of $\alpha=1.2$ and $1.5$ 
    are out of range for $t>100$ in (b) and (c).
    (d) The horizontal axis is scaled as $t/N^{\beta}$.
    (inset) $\alpha$ dependence of $\beta$,
    which is obtained by fixing $\beta=1$ for $(N,\alpha)=(8192,0)$.    
    The lower red and upper blue lines are for $N=1024$ and $N=8192$
    respectively.
    The black solid line represents $\beta=1-\alpha^{2}$.}
  \label{fig:alphaHMF-lowFAT}
\end{figure}

Fixing $T=0.45 (<T_{\rm c})$, we computed power spectra of $M^{2}$
by varying the value of $\alpha$ (Fig. \ref{fig:alphaHMF-low}).
Note that, for $\alpha>1$, the initial condition randomly chosen from
the distribution function $A \exp \left[ -(p^{2}/2-M\cos q)/T \right]$
is not guaranteed to be in thermal equilibrium.
As in the high-energy disordered phase,
the slope of power spectrum goes to zero as $\alpha$ increases,
but $\alpha=1$ seems not the threshold.
Instead, the slope goes to zero at $\alpha$
beyond which the ordered phase disappears.
We may imagine that, in the low-energy side,
wrecks of the Casimir invariants remain
for $\alpha\gtrsim 1$.
This expectation does not contradict with the fact
that the Casimir invariants exist if the interaction is long-range,
since validity of the converse is not mentioned.
One difficulty studying around $\alpha=1$ is that
$N$ dependence becomes week as shown in Fig.~\ref{fig:alphaHMF-lowFAT}.
An in-depth investigation in
the middle range of $\alpha\gtrsim 1$
needs to be performed.

\section{Dependence of observables}
\label{sec:dep-observables}

In the globally coupled Fermi-Pasta-Ulam-Tsingou system
\begin{equation}
  \label{eq:FPUT}
  H_{\rm FPUT} = \sum_{j=0}^{N-1} \dfrac{p_{j}^{2}}{2}
  + \dfrac{1}{2N} \sum_{j=0}^{N-1} \sum_{k=0}^{N-1} \varphi(q_{j}-q_{k}),
  \quad
  \varphi(q) = \dfrac{q^{2}}{2} + \dfrac{q^{4}}{4},
\end{equation}
we compute the other power spectra of the variance of $q$,
\begin{equation}
  \Delta = X_{2} - X_{1}^{2},
  \quad
  X_{n} = \dfrac{1}{N} \sum_{j=0}^{N-1} q_{j}^{n},
\end{equation}
the potential energy $U$, and $\boldsymbol{\Theta}^{2}$,
where the order parameter $\boldsymbol{\Theta}$ is defined by
\begin{equation}
  \boldsymbol{\Theta} = \dfrac{1}{N} \sum_{j=0}^{N-1}
  (\cos\theta_{j}, \sin\theta_{j}),
\end{equation}
and $\theta$ is the angle variable, which is conjugate
to the action variable associated with the one-particle Hamiltonian
\begin{equation}
  \mathcal{H}_{\rm FPUT} = \dfrac{p^{2}}{2} + \dfrac{1+3X_{2}^{c}}{2} q^{2} + \dfrac{1}{4}q^{4}.
\end{equation}
The value of $X_{2}^{c}$ is computed to satisfy the self-consistent equation
from the canonical thermal equilibrium distribution,
which is proportional to $\exp(-\mathcal{H}_{\rm FPUT}/T)$.
For $T=1$, we have $X_{2}^{c}\simeq 0.34252$.

\begin{figure}[h]
  \centering
  \includegraphics[width=8cm]{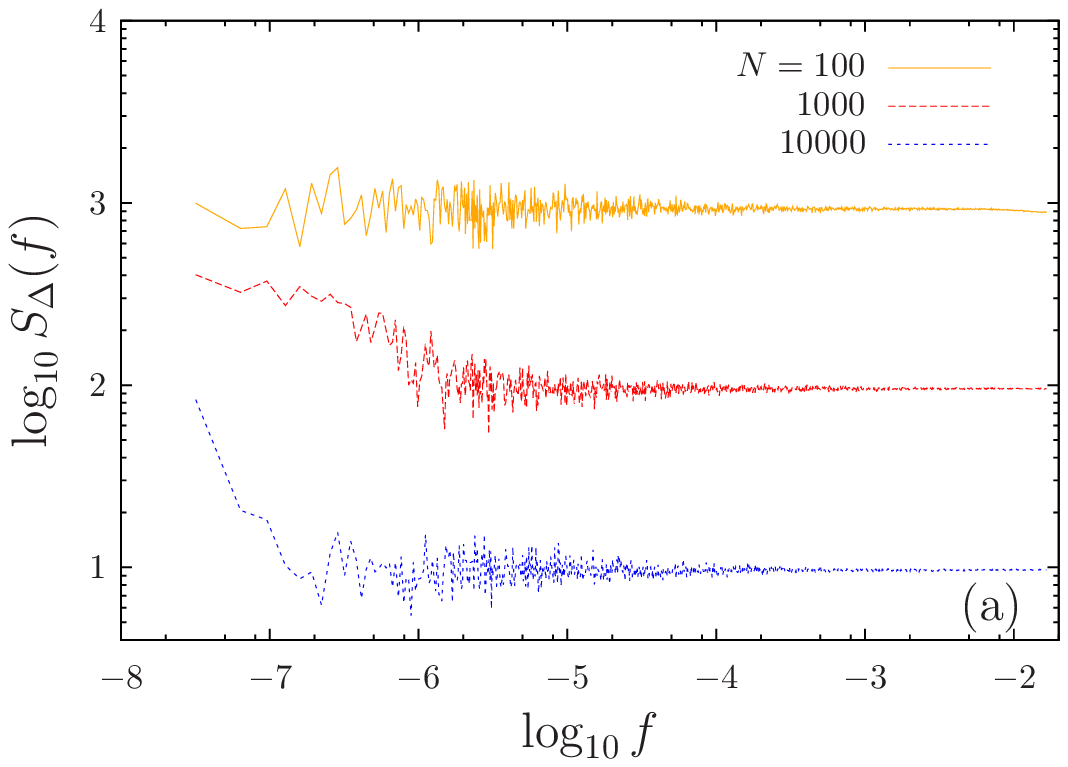}
  \includegraphics[width=8cm]{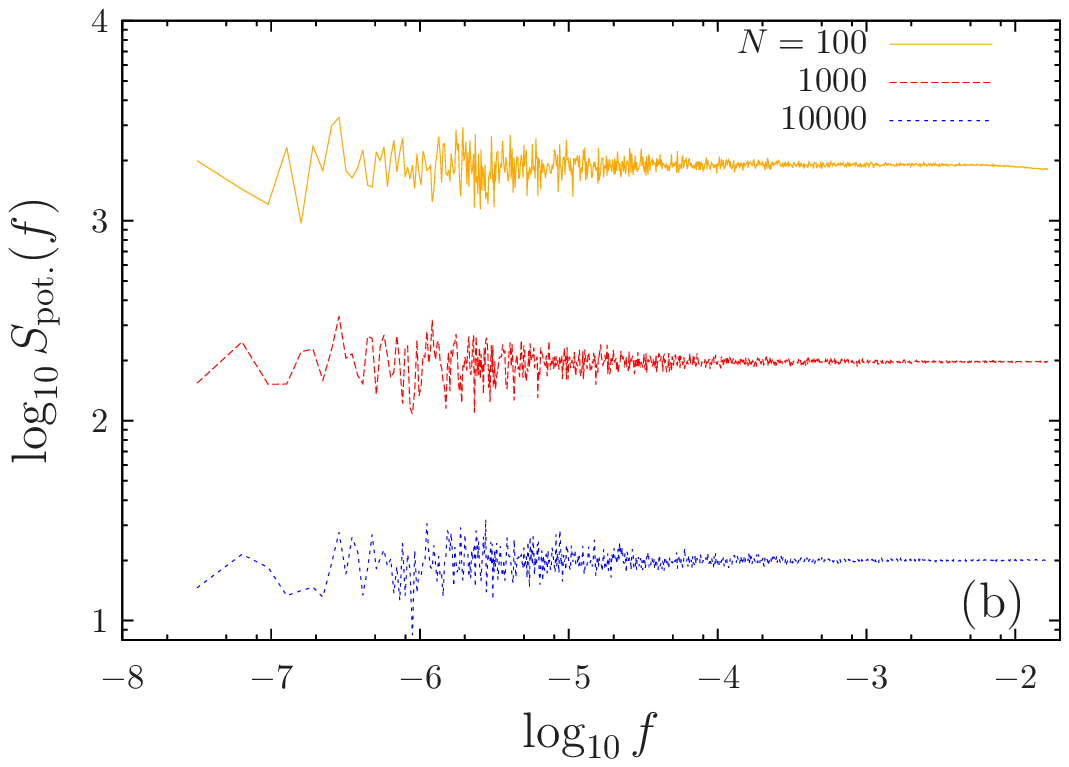}
  \includegraphics[width=8cm]{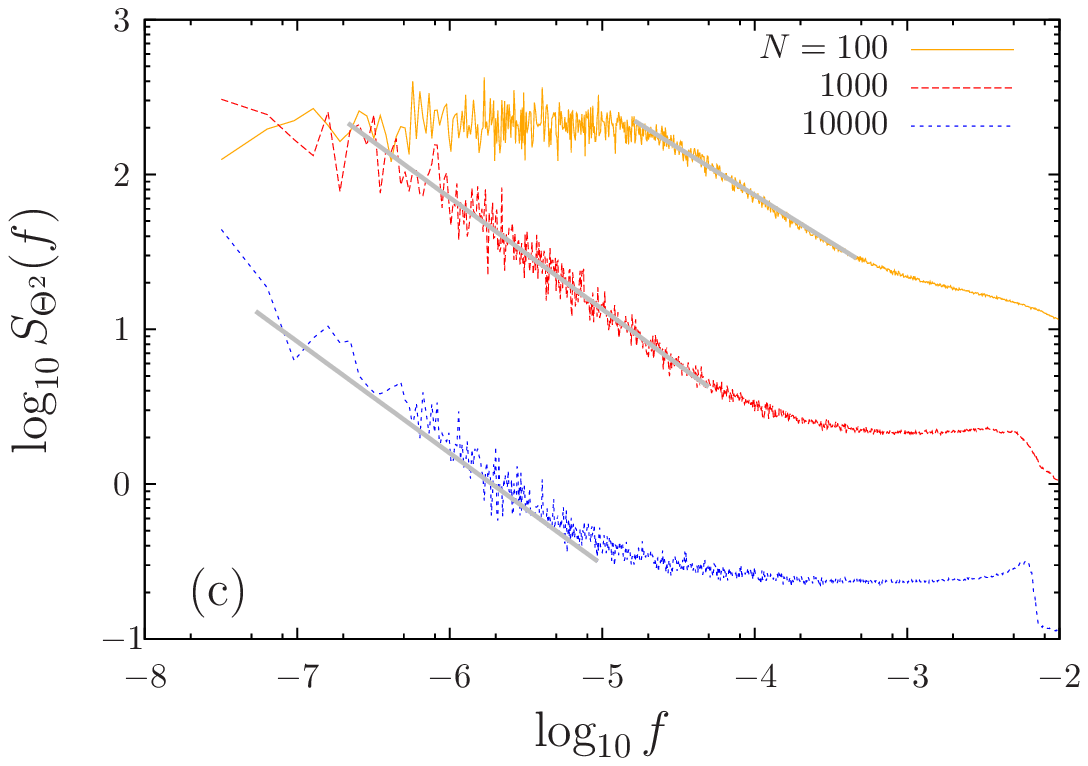}
  \caption{(color online)
    Power spectra of (a) $\Delta$, (b) potential energy,
    and (c) $\boldsymbol{\Theta}^{2}$
    in the globally coupled FPUT model. $T=1$.
    $N=100$ (top orange), $1000$ (middle red), and $10000$ (bottom blue),
    where the vertical scales of the last two lines in (c)
    have been multiplied by $10$ and $100$, respectively, for graphical reasons.
    The curves are averages over $20$ samples
    obtained with the time step of $\Delta t=0.01$.
    The gray line segments in (c) are obtained by the least square method
    in the intervals of the segments,
    and their slopes are $-0.61$, $-0.72$, and $-0.72$ for $N=100$, $1000$,
    and $10000$ respectively.}
  \label{fig:FPUT-others}
\end{figure}

The power spectra are presented in Fig.~\ref{fig:FPUT-others}.
The variables $\Delta$ and $U$ have no $1/f$ fluctuation,
but $\boldsymbol{\Theta}^{2}$ is another suitable observable
to find the hidden $1/f$ fluctuation as $\boldsymbol{\Phi}^{2}$.

\section{Kuramoto model}
\label{sec:kuramoto-model}

The Kuramoto model is expressed as
\begin{equation}
  \label{eq:Kuramoto-model}
  \dfrac{d\theta_{j}}{dt} = \omega_{j} - \dfrac{K}{N} \sum_{k=1}^{N} \sin(\theta_{j}-\theta_{k}), 
\end{equation}
where $\theta_{j}$ represents the phase of the $j$th oscillator,
$\omega_{j}$ is the natural frequency depending on the number of oscillators,
and $K>0$ is the coupling constant.
The natural frequency $\omega_{j}$ is independently and randomly
drawn from a probability distribution function $g(\omega)$.
The Kuramoto model describes the synchronization transition
from the nonsynchronized state to the partially synchronized state.
The extent of synchrony is measured by the order parameter
\begin{equation}
  Z = \dfrac{1}{N} \sum_{k=1}^{N} e^{i\theta_{k}}.
\end{equation}
If $g(\omega)$ is even and unimodal, then we find the critical
coupling constant $K_{\rm c}=2/[\pi g(0)]$,
where the nonsynchronized state is stable (unstable)
for $K<K_{\rm c}$ $(K>K_{\rm c})$.

In the limit $N\to\infty$,
the Kuramoto model is described by the equation of continuity
\begin{equation}
  \label{eq:eq-continuity}
  \dfrac{\partial f}{\partial t} + \dfracp{}{\theta}( vf) = 0,
\end{equation}
where $f(\theta,\omega,t)$ is the probability distribution function
satisfying $\int_{0}^{2\pi} f(\theta,\omega,t) d\theta = g(\omega)$, and
\begin{equation}
  v = \omega - K \int_{-\infty}^{\infty} d\omega \int_{0}^{2\pi} d\theta' ~ \sin(\theta-\theta') f(\theta',\omega,t).
\end{equation}
The equation of continuity \eqref{eq:eq-continuity}
corresponds to the Vlasov equation \eqref{eq:Vlasov-eq}.
This correspondence becomes clear by rewriting the Vlasov equation
in the form
\begin{equation}
  \dfracp{F}{t} + \nabla_{(q,p)}\cdot \left( \boldsymbol{X} F \right) = 0,
\end{equation}
where
\begin{equation}
  \nabla_{(q,p)} = \left( \dfracp{}{q}, \dfracp{}{p} \right) 
\end{equation}
and $\boldsymbol{X}$ is the Hamiltonian vector field
\begin{equation}
  \boldsymbol{X} = \left( \dfracp{\mathcal{H}[f]}{p}, -\dfracp{\mathcal{H}[f]}{q} \right).
\end{equation}
Nevertheless, the equation of continuity
has no Poisson structure and no Casimir invariants accordingly.
We can not expect $1/f$ fluctuation of collective variables
originated from the Casimir mechanism.

\begin{figure}[h]
  \centering
  \includegraphics[width=8cm]{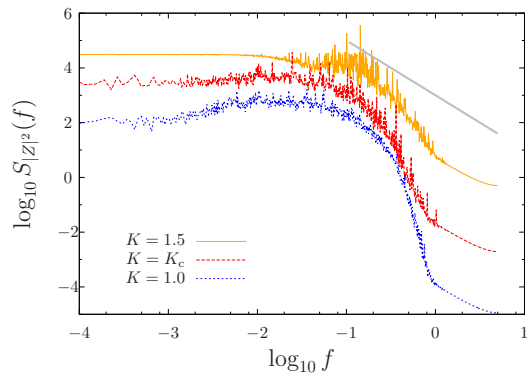}
  \caption{(color online)
    Power spectra of $|Z(t)|^{2}$ in the Kuramoto model.
    The coupling constant is $K=1.0$ (blue lower),
    $K=K_{\rm c}$ (red middle), and $K=1.5$ (orange upper).
    The vertical scale for $K=1.5$ is multiplied by $10$
    for graphical reasons.
    Each curve is the average over $20$ samples.
    The upper gray line segment guides the eyes for the slope $-2$.
  }
  \label{fig:Kuramoto}
\end{figure}

To confirm absence of $1/f$ fluctuation in the order parameter,
we set $g(\omega)$ as the normal distribution
\begin{equation}
  g(\omega) = \dfrac{1}{\sqrt{2\pi T}} \exp( -\omega^{2}/2T )
\end{equation}
with ``temperature'' $T=0.6$, which gives $K_{\rm c}=1.236...$.
We numerically integrate the equation of motion \eqref{eq:Kuramoto-model}
by using the fourth-order Runge-Kutta method
with the time step $\Delta t=0.1$.
The power spectrum of $|Z(t)|^{2}$ is reported in Fig.~\ref{fig:Kuramoto}.
The observable is a collective variable,
but no $1/f$ fluctuation appears irrespective of values of
the coupling constant $K$.

\section{Superposition of Lorentzian spectra}
\label{sec:superposition-lorentzian}

One traditional explanation of $1/f$ fluctuations
is a superposition of Lorentzian spectra originating
from the exponentially damping correlation
(see \cite{milotti-02,ward-greenwood-07}, for instance).
A long-range system has several exponential Landau damping modes in general,
but the $1/f$ spectra reported in this article cannot be explained
by this superposition:
The second slowest damping rate is of $O(1)$ around $T=T_{\rm c}$
in the $\alpha$-HMF model,
and no accumulation of Landau poles appears 
in the slow timescale of $f\sim 10^{-3}$.

\twocolumngrid


\begin{thebibliography}{99}
\bibitem{comment:cf}
  In our study, $F$ rapidly goes to zero in $|p|\to\infty$
  and $c(F)=F^{n}~(n=1,2,\cdots)$ make Casimir invariants,
  for instance.
\bibitem{comment:kappa-alpha}
  We note that the divergence of $\kappa_{\alpha}$ can be avoided
  by introducing, for instance, long-range-like coupling constants
  in the interval $x\in [-\epsilon,\epsilon]$ with small $\epsilon>0$.
  However, the scaling $x=j/N$ says that the particle at $x=0$ interacts
  with an infinite number of particles in a long-range manner,
  and the short-range nature disappears accordingly.
  The case $\alpha=1$ is included in the long-range coupling,
  but we do not discuss this boundary case.
\bibitem{comment:FFT}
  For $N=2^{n},~n\in\mathbb{N}$,
  the interaction coefficient matrix, $R=(1/(N_{\alpha}r_{jk}^{\alpha}))$,
  is diagonalized by the fast Fourier transform (FFT) matrix.
  By using this diagonalization and the FFT algorithm, 
  we can reduce the numerical cost of the $\alpha$-HMF model
  from $O(N^{2})$ to $O(N\log N)$.
  This trick helps to perform long-time evolution.
\bibitem{comment:Vt-initial}
  The initial increase in $V(t)$ for $t<10$ comes from the
  definition \eqref{eq:Vt}, which gives $V(1)=0$.
\bibitem{comment:time-shift}
  The constraint by the pseudo-Casimir-invariants remains,
  even after the variance reaches the asymptotic level.
  Thus, a similar evolution of $V(t)$ is obtained
  against the shift of the initial time \cite{yamaguchi-16}.
\bibitem{comment:InitialTransience}
  No transience has been observed in the HMF model \cite{yamaguchi-16},
  so this procedure was skipped in the HMF model.
\bibitem{comment:action-angle}
  The variable $\phi$ does not completely agree with the angle variable
  which should be obtained with the action variable 
  by the canonical transform of $(q,p)$ \cite{goldstein-poole-safko-01},
  but $\phi$ approximately represents the angular motion.
\end{thebibliography}
\end{document}